# An observation of a new class of afterpulses with delay time in the range of 70-200 μs in classical vacuum photomultipliers.

R.V. Poleshchuk[1], B.K. Lubsandorzhiev[1,2*], R.V. Vasiliev[1]

[1]*Institute for Nuclear Research of the Russian Academy of Sciences, Moscow, Russia*

[2]*Kepler Centre for Astro and Particle Physics, University of Tübingen, Germany*

∗ *Corresponding author: postal address: pr-t 60th Anniversary of October, 7a, 117312 Moscow, Russia; phone: +7-499-1353161; fax: +7-499-1352268;*

E-Mail: lubsand@rambler.ru, lubsand@pit.physik.uni-tuebingen.de

**Abstract**

We present results of the first observation of afterpulses with extremely long (~120 μs) delay time from the main pulse in 8" classical vacuum photomultipliers.

## 1. Introduction

Despite the fact that afterpulses in classical photomultipliers have been known for many decades [1-9] they are still not very well studied and it continues to be intensely debated. The issue is of utmost importance in the light of upcoming next generation astroparticle physics experiments like LENA [10] putting very strong requirements on the rate of afterpulses in vacuum photodetectors.

On the other hand, a fundamental problem correlated with the increasing sensitivity of PMTs is the also increasing of afterpulse rate [11].

## 2. Afterpulses in photomultipliers

There are several models explaining the origin of afterpulses. One can subdivide afterpulses into two groups: fast and long delayed afterpulses [8]. Further in this paper we will call them "fast" and "long" afterpulses. The fast afterpulses occur within 100 ns after the main pulses, whereas the long ones come within 100 ns -20μs.

The fast afterpulses can be explained quite satisfactorily by the light feedback: photons produced in ionization of atoms and molecules of residual gas and cathode-luminescence of dynode surface; photons produced by deceleration processes (X-ray) and transitional radiation emission.

It is worth to mention here that one should distinguish between fast afterpulses and late pulses. Late pulses are in fact the main pulses but only delayed due to photoelectron backscattering on the first dynode [12, 13].

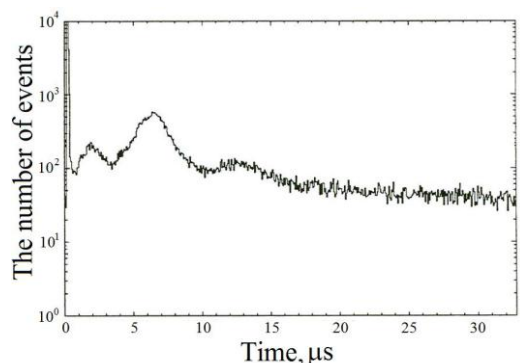

Fig. 1. Typical time distribution of long afterpulses in 8" photomultiplier Thorn-EMI9350KB.

The long afterpulses are due mostly to ionic feedback – ions produced by ionization of residual gas atoms and molecules in the PMT

volume and on the dynode surfaces are accelerated back to photocathode and dynodes and yield secondary electrons. A typical time distribution of the long afterpulses for 8" photomultiplier Thorn-EMI9350KB from the TUNKA experiment is shown in Fig.1. The afterpulses extend up to 15 μs further merging together with random coincidences due to PMT's spurious dark current pulses in the larger delay time range. In Fig.2 delay time distributions of afterpulses measured with 6" PMT FEU-49B in the wide range of 0-180 μs with (black) and without (red) LED pulses are shown. All measurements of afterpulses delay time distributions described above have been done with "Start-Stop" method using conventional wide range TDC [14] and fast LED driver [15, 16] providing light pulses with $\lambda_{max}$=470 nm and ~2 ns width (FWHM) at ~1 KHz repetition rate. The method implies that the afterpulse measurement is distorted by the underlying background from the dark count rate of the PMT described by a distribution of random time intervals:

$$W(t) \sim N_0 \exp(-N_0 t) \qquad (1)$$

where $N_0$ – mean dark current counting rate of PMT [17].

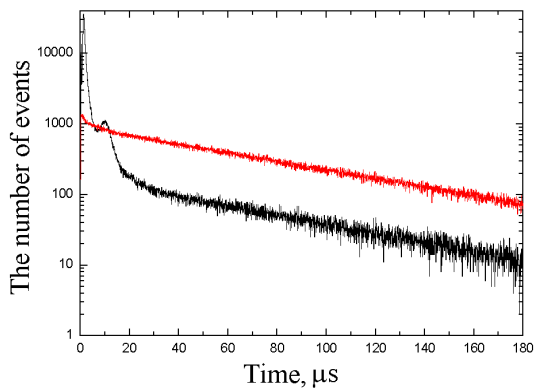

Fig. 2. Typical time distributions of long afterpulses (black curve) and random coincidences due to dark current pulses (red curve) in 6" photomultiplier FEU-49B.

So far the longest observed afterpulses did not extend for more than 40 μs [4, 7]. Zhao proposed to explain the origin of afterpulses with delay time up to 40 μs by glass fluorescence. Ions produced by electrons in the dynode system are deflected to hit PMT's glass bulb causing glass fluorescence. However, the time range of 40 μs in this model is reached only by fluorescence exponential tail without any peaks in distribution.

In the early 1980s Glukhovskoy and .Yaroshenko [18] applied exoelectronic models [19] along with ionic feedback models to explain production of afterpulses in PMT with delay time up to 20 μs.

Moreover there are still a lot of questions concerning which ions ($H^+$, $H_2^+$, $He^+$, $H_2O^+$, $CH_4^+$, $Cs^+$, etc.) and where exactly in PMT do they originate from? But nevertheless this class of afterpulses is more or less well studied and understood.

### 3. New class of afterpulses?

Testing PMTs for the TUNKA and Double-Chooz experiments we observed, quite unexpectedly, afterpulses in two samples of 8" PMTs produced by Electron Tubes and Photonis with delay time in the range of 70-200 μs from the main pulses. The delay time distribution of such afterpulses for one sample of Thorn-EMI9350KB PMT is presented in Fig. 3 (red curve). For comparison the distribution for FEU-49B PMT (black curve) is shown in the same plot.

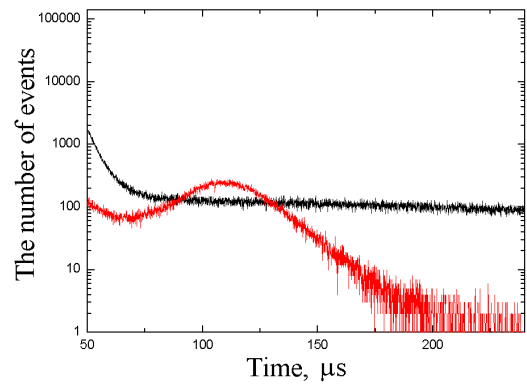

Fig. 3. Time distribution of extremely long afterpulses in 8" photomultiplier Thorn-EMI9350KB (red curve). Black curve is time distribution of afterpulses for 6" photomultiplier FEU-49B.

For Thorn-EMI9350BK PMT there is a pronounced peak with the mean delay value of 120 μs. Whereas for FEU-49B there is no such peak. Once again, both distributions were measured by conventional TDC. So, the real afterpulses delay time distributions are distorted by exponential tail of random coincidences in accordance with the expression (1). In order to avoid such distortion we measured the afterpulses delay time distributions for the same samples of PMTs using FADC registering signals waveforms in the time range of 0-1 ms. The results of FADC measurements are shown in Fig. 4. The PMT was illuminated by short light pulses form LED with intensity corresponding to 500 photoelectrons. In the left part of the distributions there exist ordinary long afterpulses with peaks around 1 and 10 µs.

The afterpulses delay time distributions for 500 pe and 20 pe illuminations are shown in Fig. 5, black and green plots, respectively. In case of 500 pe illumination the peak due to extremely long delayed afterpulses is distinctly observed in the time range of 70-200µs with mean delay time of 120 µs. In the second case the peak is not so well pronounced. Furthermore there are no peaks of afterpulses up to 1 ms.

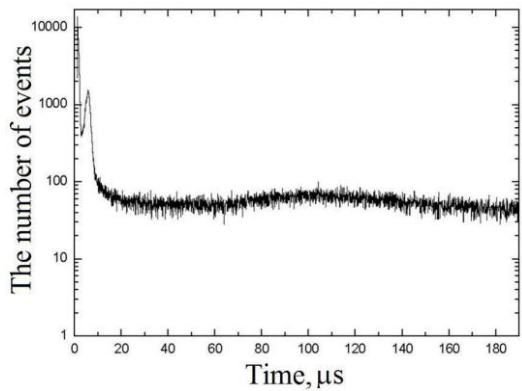

Fig. 4. Time distribution of extremely long afterpulses in 8" photomultiplier Thorn-EMI9350KB. The amplitude of the main pulse – ~500 photoelectrons.

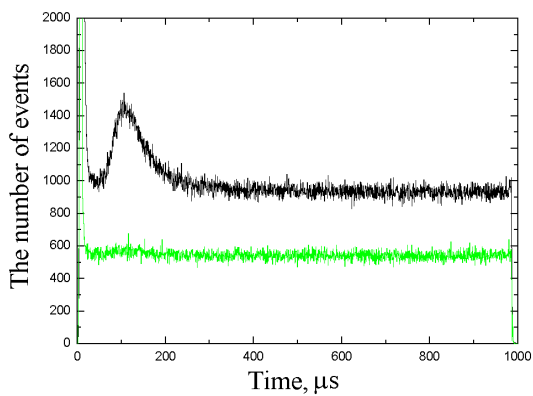

Fig. 5. Time distribution of extremely long afterpulses in 8" photomultiplier Thorn-EMI9350KB in the time range of 0 – 1 ms. The amplitude of the main pulse – ~500 pe (black curve) and ~20 pe (green curve).

It should be noted that amplitudes of these extremely long delayed afterpulses stay strongly at a single photoelectron level. The rate of the afterpulses is rather low. The probability of their production is much less than 0.1% per one photoelectron.

## 4. Conclusion

A new class of afterpulses with extremely long delay time in the range of 70-200 µs from the main pulse has been observed. So far, the afterpulses have been observed only in two samples of 8" photomultipliers of two types. The origin of the observed afterpulses is not understood, but it is quite clear that they are well out of scope of any presently existing models. Despite the very low rate, their existence may put restrictions on designing of future experiments with photomultipliers, e.g. experiments searching for massive slow moving objects, and based on time-of-flight technique.

The authors are indebted very much to Dr. V. Ch. Lubsandorzhieva for many invaluable discussions and remarks which were indispensable for completing the paper.